\documentclass[12pt]{article}
\usepackage{graphicx}
\title{EARTH: 15 MILLION YEARS AGO}
\date{}
\author{{ Masataka Mizushima}\\ Department of Physics, University of Colorado\\
Boulder, Colorado 80309}
\begin{document}
\date{}
\maketitle{}
\date{}
\begin{abstract}

In Einstein's general relativity theory the metric component $g_{xx}$ in
the direction of motion
(x-direction) of the sun deviates from unity due to a tensor potential caused
by the black hole existing around the center of the galaxy. Because the solar system is orbiting
around the galactic center at 200 km/s, the theory
shows that the Newtonian gravitational potential due to the sun is not quite
radial. At the present time, the ecliptic plane is almost perpendicular to the
galactic plane, consistent with this modification of the Newtonian
gravitational force. The ecliptic plane is assumed
to maintain this orientation in the galactic space as it orbits around the galactic
center, but the rotational angular momentum of the earth around its own axis can be
assumed to be conserved. The earth is between the sun and the galactic center
at the summer solstice all the time.
As a consequence, the rotational axis of the earth would be parallel to the axis
of the orbital rotation of the earth 15 million years ago, if the solar system
has been orbiting around
the galactic center at 200 km/s. The present theory concludes that the earth did
not have seasons 15 million years ago. Therefore, the water on the earth was accumulated
near the poles as ice and the sea level was very low. Geological
evidence exist that confirms this effect. The resulting global ice-melting started 15 million years ago and is ending now.\\

{\em Subect heading}:  General Relativity, Solar System, Milky Way Galaxy
\end{abstract}

\pagebreak

If we take
\begin{equation}
\renewcommand{\theequation}{1}
g_{ij} = g_{ij}^{(\circ)} + h_{ij},
\end{equation}
where
\begin{equation}
\renewcommand{\theequation}{2}
g_{00}^{(\circ)} = 1,~~g_{\alpha 0}^{(\circ)} = 0,~~g_{\alpha\beta}^{(\circ)} = -\delta_{\alpha\beta},
\end{equation}
 $\alpha$ = 1, 2, 3, for the general relativistic metric
\begin{equation}
\renewcommand{\theequation}{3}
(ds)^{2} = g_{ij}dx^{i}dx^{j},
\end{equation}
and assume that the $h_{ij}$ are all so small that their powers higher than the first
are negligible, Einstein's general relativistic field equations can be
simplified into those of the linearized Einstein theory.

Introducing
\begin{equation}
\renewcommand{\theequation}{4}
h_{ij} = \phi_{ij} - \frac{1}{2}\phi g_{ij}^{(\circ)}
\end{equation}
where $\phi = \phi^{i}_{i}$, and $\phi^{i}_{j} = g^{(\circ)ik}\phi_{kj}$,
and assuming that
\begin{equation}
\renewcommand{\theequation}{5}
\frac{\partial\phi^{i}_{j}}{\partial x^{i}} = 0,
\end{equation}
we see that Einstein's field equation can be approximated as
\begin{equation}
\renewcommand{\theequation}{6}
\left( \frac{\partial^{2}}{\partial x^{2}}
- \frac{1}{c^{2}}\frac{\partial^{2}}{\partial t^{2}}\right)\phi^{i}_{j}
= \frac{16\pi G}{c^{4}}T^{i}_{j}.
\end{equation}
In eq. (6), $T^{i}_{j}$ is a component of the mass energy-momentum
tensor such that $T^{0}_{0} = \mu c^{2}$ .

It is easy to solve eq. (6), because it is mathematically identical to
the Maxwell field equations, to obtain
\begin{equation}
\renewcommand{\theequation}{7a}
\phi_{00} = - \Sigma_{a} \frac{4GM_{a}}{c^{2}r_{a}}\mid_{t'},
\end{equation}
\begin{equation}
\renewcommand{\theequation}{7b}
\phi_{\alpha 0} = \Sigma_{a} \frac{4GM_{a}v_{a,\alpha}}{c^{3}r_{a}}\mid_{t'},
\end{equation}
\begin{equation}
\renewcommand{\theequation}{7c}
\phi_{\alpha\beta} = - \Sigma_{a} \frac{4GM_{a}v_{a,\alpha}v_{a,\beta}}{c^{4}r_{a}}\mid_{t'},
\end{equation}
where $M_{a}$ is the mass of a point source located at a distance $r_{a}$ from the observation
point, and $v_{a,\alpha}$ is the $\alpha$-component of its velocity.
Because eq. (6) is linear in $\mu$, the mass density, we see that the
right-hand sides of eqs, (7a) through (7c) are given by the sum over existing point masses $M_{a}$.
Quantities on the right-hand side of eqs. (7a) through (7c) are to be evaluated at
the retardation time $t'$, but the retardation effect can be neglected. \\
\indent In the following discussion, we take the solar mass as $M_{s}$ and the
black hole mass as $M_{b}$, which exists near the center of the Milky way galaxy. In the Newtonian
limit, $v/c \rightarrow$ 0, we see that
\begin{equation}
\renewcommand{\theequation}{8}
g_{00} = -1 + \frac{1}{2}\phi_{00} = -1 - \frac{2GM_{s}}{c^{2}r_{s}}
- \frac{2GM_{b}}{c^{2}r_{b}}
\end{equation}
where $r_{s}$ and $r_{b}$ are the distances to the sun and to the galactic
center, respectively. The terms proportional to $M_{s}$ and $M_{b}$ give the
Newtonian potentials due to the sun and the black hole at the galactic center,
respectively. The Newtonian potential due to $M_{b}$ is negligible compared to that due to the sun.
In the solar system, the solar mass $M_{s}$ is not moving,
but the source mass $M_{b}$ is moving in the direction tangential to the
galactic center, which we call the $x$-direction, so that
$\phi_{yy} = \phi_{zz} = 0$, and
\begin{equation}
\renewcommand{\theequation}{9}
g_{xx} - g_{yy} = g_{xx} - g_{zz} = \frac{1}{2}\phi_{xx}
= -\frac{2GM_{b}v^{2}_{g}}{c^{4}r_{b}} \equiv -P,
\end{equation}
where $v_{g}$ is the velocity of the galactic center with respect to the
sun, in the direction tangential to the galactic center. Actually, we can
write this anisotropy $P$ as
\begin{equation}
\renewcommand{\theequation}{10}
P = \frac{2GM_{b}v^{2}_{g}}{c^{4}r_{b}} = 2(v_{g}/c)^{4},
\end{equation}
using the relation $GM_{b}/r_{b} = v^{2}_{g}$, for a circular orbit. Because the orbiting
speed $v_{g}$ is 200 km/s, the anisotropy is $P = 2\times 10^{-12}$ at the solar
system. The retardation effect is neglected in this estimation.

Einstein also proposed an equation of motion under gravity:
\begin{equation}
\renewcommand{\theequation}{11}
\frac{d^{2}x^{i}}{ds^{2}} = - \Gamma^{i}_{jk}\frac{dx^{j}}{ds}\frac{dx^{k}}{ds},
\end{equation}
where
\begin{equation}
\renewcommand{\theequation}{12}
\Gamma^{i}_{jk} = \frac{1}{2}g^{im}\left(\frac{\partial g_{mk}}{\partial x^{j}}
+ \frac{\partial g_{mj}}{\partial x^{k}} - \frac{\partial g_{kj}}{\partial x^{m}}\right)
\end{equation}
is a component of the Christoffel symbol. In the nonrelativistic limit,
$s = ct$, and the first term is obtained by taking $j = k = 0$ in eq. (11),
so that
\begin{equation}
\renewcommand{\theequation}{13}
\frac{d^{2}x}{dt^{2}} \approx -\Gamma^{1}_{00}
= \frac{c^{2}}{2}g^{xx}\frac{\partial g_{00}}{\partial x}.
\end{equation}
To the first order in $P$, we obtain
\begin{equation}
\renewcommand{\theequation}{14}
g^{xx} = 1 - \frac{1}{2} \phi_{xx} = 1 + P,
~~g^{yy} = g^{zz} = 1 .
\end{equation}
neglecting $\phi_{00}$ compared to 1. Therefore, eq. (13) reduces to
\begin{equation}
\renewcommand{\theequation}{15}
\frac{d^{2}{\bf r}}{dt^{2}} = - ({\bf\nabla} +
(P{\bf v}_{g}/v^{2}_{g})({\bf v}_{g}\cdot{\bf\nabla}))\frac{GM_{s}}{r_{s}},
\end{equation}
where $M_{s}$ is the solar mass. In eq. (7a), $\phi_{00}$ is the sum of
the Newtonian potentials due to the galactic center
and to the sun, but the contribution to eq. (15) due to the galactic
center is negligible. The extra term, proportional to $P$, in eq. (15),
provide the torque to keep the ecliptic plane  containing the galactic
center during the orbital motion of the solar system around the galactic
center (Mizushima., 1992)

\begin{center}
\begin{figure}
\includegraphics[width=10cm]{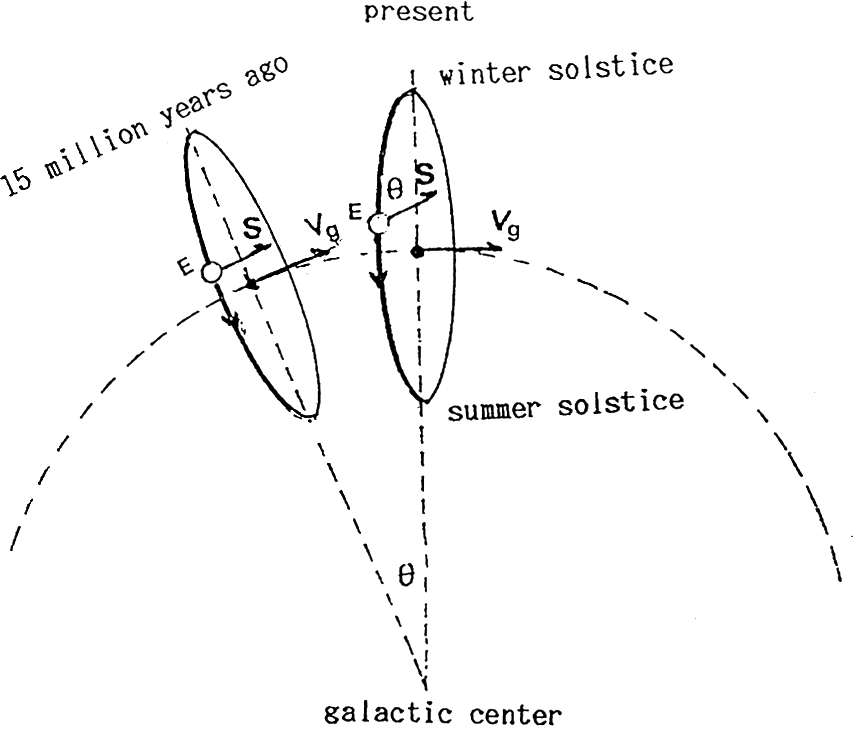}
 \caption{At present the ecliptic plane is almost perpendicular to the orbiting
 velociy ${\bf v}_{g}$ of the solar system around the galactic center; the
 ecliptic plane contains the galactic center in such a way that the earth is
 between the galactic center and the sun at the summer solstice, which implies that the
 rotional angular momentum of the earth, ${\bf S}$, is in the galactic plane, about
 $\theta = 23.4^{\circ}$ with respect to ${\bf v}_{g}$ at present. Assuming that the
 orbiting speed, 200 km/s, remained constant with the radial distance
 $3\times 10^{20}$ m from the galactic center, orbiting velocity ${\bf v}_{g}$
 was $23.4^{\circ}$ off from the present direction 15 million
 years ago. Therefore, ${\bf v}_{g}$ was parallel to ${\bf S}$ 15 million
 years ago.}
\end{figure}
\end{center}

\indent Because the solar system is orbiting around the galactic center, the
Newtonian gravitational force due to the sun is not isotropic, but
anisotopic in the direction of the orbiting velocity around the galactic
center. The present orientation of the ecliptic plane is almost in the
direction that contains the galactic center, as shown in Fig.~1, and that
can be understood taking this general relativistic anisotropy of the
Newtonian gravitational force into account.\\
\indent According to the present theory, the anisotropy of the space force
kept the orientation of the ecliptic plane perpendicular to the ${\bf v}_{g}$
vector. Assuming that
the orbiting speed has been $200$ km/s, the vector ${\bf r}_{b}$ $15$ million
years ago was off by $23.4^{\circ}$ from the present direction of ${\bf r}_{b}$.
Another interesting situation of the present orbit of the earth is that the sun
is between Earth and the galactic center at the winter solstice, and the earth
is between the sun and the galactic center at the summer solstice. Thus,
the rotational angular momentum of the earth is in the galactic plane making an
angle of $23.4^{\circ}$ away from the galactic center. Therefore, $15$
million years ago, the orbital angular momentum of the earth around the sun was
parallel to its rotational angular momentum ${\bf S}$ (Fig.1).\\
\indent At present, the sun is at 23.4$^{\circ}$ North at the summer solstice
and 23.4$^{\circ}$ South at the winter solstice. Analyzing the reported average temperature at a given
location on Earth, we see that the freezing temperature at that location
appears when the angular distance of the ecliptic plane from the vertical of
that location is more than 50$^{\circ}$ at sea level and 40$^{\circ}$ in the
high mountains. The average temperature at each location is mostly dependent
on the rate of energy input by the solar radiation, which is proportional to
$\cos^{2}\theta$ where $\theta$ is the angular distance between the sun at
noon and a vertical at the observational location on the earth. (Mizushima, 2008a).\\
\indent If we assume that solar radiation was the main source of the variable
energy input to the surface of the earth 15 million years ago, as it is now, the above
observation means that the earth everywhere with a latitude greater than
40$^{\circ}$, both North and South, stayed at a temperature below freezing
year around 15 million years ago because there were no seasons. The equatorial
region stayed warm, and water kept evaporating from the oceans. But atmospheric
water vapor fell down to the surface of the earth as snow at higher latitude
regions both North and South. On the continents, the snow did not melt, and
accumulated as ice. Although ice also evaporates, it does so
very slowly. Thus, a considerable fraction of earth's water remained as ice during
the period around 15 million years ago, which resulted in very low sea
level. The geological evidence to confirm this effect exists (Vail, et al., 1977), as shown
in Fig.~2.

\begin{center}
\begin{figure}
  \includegraphics[width=10cm]{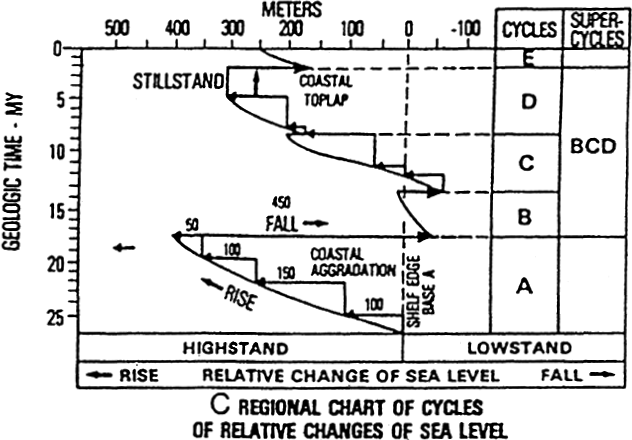}
  \caption{Geological evidence that sea level
  was at its lowest 15 million years ago (Veil, et al., 1982).}
\end{figure}
\end{center}

\indent As the ecliptic plane began to deviate from the equator, temperatures at
critical latitudes, both North and South, rose, and ice started to melt
in the summer seasons, and will keep melting for some time to come. \\ 
It is found (Mizushima, 2008b) that the tangential velocity of the solar system around the galactic center, the orbiting speed $v_{g}$ as cited in eq. (9) and Fig. 1, was very much smaller that the present value 200 km/s during the earlier life of the solar system. If that is true, the anisotropy factor P was much smaller than its present value. Therefore, the anisotropy effect shown by eq. (15) appeared only during the very recent period as discussed here. \\ \\

\section*{References}
Mizushima, M., 1992,  J. Phys. Soc. Japan {\bf 61}, 2673  \newline
Mizushima, M., 2008a, Proc. Dynamic Sys. Appl. {\bf 5}, 339 \newline 
Mizushima, M. 2008b, arXiv 0801,4809 \newline
Vail, P. R., Mitchem ,R. M. Jr., and Thomson, S. III, 1977,
	{\em Seismic Stratigraphy. Application to Hydrocarbon Exploration,
        Memoir 26} p.85, reproduced on p.278 of {\em Marine Geology},
	ed. Kennett, J. P. Prentice Hall, Englewood, NJ (1982).

\end{document}